\documentclass[english,12pt,sort&compress]{iopart}
\pdfoutput=1

\usepackage{iopams}
\usepackage{setstack}

\usepackage{psfrag}

\usepackage{dsfont} % for unit operator  (indicator function) \mathds{1}
\usepackage[usenames,dvipsnames]{color}
\usepackage{verbatim}
\usepackage[numbers,merge]{natbib}

\usepackage{hyperref}
\hypersetup{ colorlinks=true, linktoc=all, linkcolor=blue, }

\newcommand{\eqn}{Eq.\ }
\newcommand{\eqns}{Eqs.\ }

\newcommand{\nutilde}{\tilde{\nu}}
\newcommand{\rk}[1]{r_{#1}}

\newcommand{\mytext}[1]{{\mbox{#1}}}
\newcommand{\mysubtext}[1]{{\mathrm{#1}}}

\usepackage{graphicx}
\newcommand{\be}{\begin{equation}}
\newcommand{\ee}{\end{equation}}
\newcommand{\bel}[1]{\begin{equation}\label{#1}}
\newcommand{\bea}{\begin{eqnarray}}
\newcommand{\eea}{\end{eqnarray}}
\newcommand{\balign}{\begin{align}}
\newcommand{\ealign}{\end{align}}
\newcommand{\ba}{\begin{array}}
\newcommand{\ea}{\end{array}}
\newcommand{\bfig}{\begin{figure}}
\newcommand{\efig}{\end{figure}}

\newcommand{\bra}[1]{\mbox{$\langle \, {#1}\, |$}}
\newcommand{\ket}[1]{\mbox{$| \, {#1}\, \rangle$}}

\newcommand{\inprod}[2]{\mbox{$\langle \, {#1} \, | \, {#2} \, \rangle$}}

\newcommand{\Prob}[1]{\mbox{${\rm Prob}\left[ \, {#1}\, \right]$}}

\newcommand{\bfn}{\mathbf{n}}
\newcommand{\bfS}{\mathbf{S}}

\begin{document}

\title[ZRP conditioned on an atypical current]{Density profiles, dynamics, and condensation in the ZRP conditioned on an atypical current}
\author{Ori Hirschberg$^{1,2}$, David Mukamel$^1$, Gunter M. Sch\"utz$^{3,4}$}
\address{$^1$ Department of Physics of Complex
Systems, Weizmann Institute of Science, Rehovot 76100, Israel}
\address{$^2$ Department of Physics, Technion, 3200003 Haifa,
Israel}
\address{$^3$ Institute of Complex Systems II, Theoretical Soft
Matter and Biophysics, Forschungszentrum J\"ulich, 52425 J\"ulich,
Germany}
\address{$^4$ Interdisziplin\"ares Zentrum f\"ur komplexe
Systeme, Universit\"at Bonn, Br\"uhler Str. 7, 53119 Bonn, Germany}

\ead{\mailto{ori.hirschberg@ph.technion.ac.il},
\mailto{david.mukamel@weizmann.ac.il},
\mailto{g.schuetz@fz-juelich.de}}

\begin{abstract}
We study the asymmetric zero-range process (ZRP) with $L$ sites and
open boundaries, conditioned to carry an atypical current. Using a
generalized Doob $h$-transform we compute explicitly the transition
rates of an effective process for which the conditioned dynamics are
typical. This effective process is a zero-range process with
renormalized hopping rates, which are space dependent even when the
original rates are constant. This leads to non-trivial density
profiles in the steady state of the conditioned dynamics, and, under
generic conditions on the jump rates of the unconditioned ZRP, to an
intriguing supercritical bulk region where condensates can grow.
These results provide a microscopic perspective on macroscopic
fluctuation theory (MFT) for the weakly asymmetric case: It turns
out that the predictions of MFT remain valid in the non-rigorous
limit of finite asymmetry. In addition, the microscopic results
yield the correct scaling factor for the asymmetry that MFT cannot
predict.

\

\end{abstract}

\date{\today}

\maketitle

%\noindent\rule{\textwidth}{1pt}

%\tableofcontents

%\noindent\rule{\textwidth}{1pt}

\section{Introduction}
\label{s_intro}

There has been considerable progress in recent years in the
understanding of nonequilibrium, current-carrying systems
\cite{Derrida2007,EvansBlytheSolversGuide,Schadschneider2010Book,ChouEtal2011,Bert15}.
Recently, much effort has been devoted to the study of fluctuations
which result in atypical currents in such systems
\cite{Derrida2007,Bert15}, where equilibrium concepts of entropy and
free energy could be generalized to a nonequilibrium setting.
Following the seminal papers
\cite{DerridaLebowitzSpeer2002SEPldfJSP,BertiniEtal2002Long}, the
non-typical large-deviation behaviour of Markovian stochastic
lattice gas models has come under intense scrutiny in the framework
of what is now known as macroscopic fluctuation theory (MFT)
\cite{Bert15}. In particular, the asymmetric simple exclusion
process (ASEP) \cite{Ligg99,Schu01} has been studied in great detail
under the condition that the dynamics exhibits a strongly atypical
current for a very long interval of time. Among the questions of
considerable importance is the optimal density profile that realizes
such a rare large deviation. Using MFT it was found for the
\emph{weakly} asymmetric simple exclusion process, where the hopping
bias is of order $1/L$, that for an atypically low current a
dynamical phase transition occurs in the case of periodic boundary
conditions: Above a critical non-typical current the optimal
\emph{macroscopic} density profile is flat, while rare events below
that critical current are typically realized by a traveling wave
\cite{BodineauDerrida2005WeaklyAsym}. This phenomenon was
subsequently been studied numerically with Monte-Carlo simulations
\cite{Hurt11,Espi13}. Non-flat optimal profiles were obtained also
for open boundary conditions \cite{Bodi06}.

More recently, the space-time realizations of large-deviation events
in the ASEP with {\it finite} (strong) hopping bias were studied on
the \emph{microscopic} scale. This was done by conditioning the
time-integrated current in a {\it finite time interval} to attain
some atypical value. Furthermore, the full conditioned probability
distribution on a \emph{finite lattice} has been examined. For low
non-typical current, the microscopic structure of a fluctuating
traveling wave and an antishock has been identified
\cite{Simo09,Beli13a,Beli13b}. Recently duality was used to extend
the microscopic approach to show that the traveling wave may exhibit
a microscopic fine structure consisting of several shocks and
antishocks \cite{Schu15b}. These findings are consistent with the
macroscopic results for the optimal density profile
\cite{BodineauDerrida2005WeaklyAsym,Bodi06}, but provide much more
information in that the complete time-dependent measure was obtained
for the microscopic dynamics.

Using a type of Doob's $h$-transform
\cite{Mill61,Doob,Jack10,Chet13} one may also study the large time
limit on microscopic lattice scale. This transformation generates
effective dynamics which make the rare large deviation behaviour of
the original dynamics typical. For the ASEP with periodic boundary
conditions and large non-typical current, inaccessible to MFT, it
was found that a different dynamical phase transition involving a
change of dynamical universality class occurs: Instead of the
well-known generic universality class of the KPZ equation with
dynamical exponent $z=3/2$ for the typical dynamics \cite{KPZ} one
has a ballistic universality class with dynamical exponent $z=1$
where fluctuations spread much faster
\cite{PopkovEtal2010AsepEnhancedFlux,Popk11,Schu15a}. In fact, it
turns out that these results follow predictions from conformal field
theory and are thus expected to be universal \cite{Kare15}.
Moreover, using the $h$-transform technique, long-range effective
interactions which make these rare events typical were obtained
\cite{PopkovEtal2010AsepEnhancedFlux}.

In this paper, we study similar questions, regarding the emergence
of atypical currents, in another prototypical model of
nonequilibrium systems --- the zero range process (ZRP)
\cite{Spitzer1970,evanszrpreview}. The steady state of the ZRP is
exactly soluble, and thus it has often been used in studies of
out-of-equilibirum features. The ZRP is a stochastic lattice gas
model where each lattice site $i$ can be occupied by an arbitrary
number $n_i \geq 0$ of particles. Particles move randomly to
neighbouring sites with a rate $u(n_i)$ that depends only on the
occupation number of the departure site and not on the state of the
rest of the lattice. We consider a one-dimensional lattice of sites
$i=1,\ldots,L$ with open boundaries where the system exchanges
particles with external reservoirs. The hopping events may be biased
in one direction.

The precise definition of the model is given below. As an
introduction we summarize some well-known features of the ZRP. The
stationary state of the model with {\it periodic} boundary
conditions, where the total particle number is conserved, may
exhibit a condensation transition above a critical density. In this
scenario, the existence of which depends on the form of the hopping
rates $u(n)$, all sites but one are occupied by a particle number
that fluctuates around the critical density, while one randomly
selected site carries all the remaining excess particles, whose
number is of the order of the lattice size (for a review see
\cite{evanszrpreview}). On the other hand, for open boundary
conditions where the total particle number fluctuates due to random
injection and absorption of particles at the two boundaries of the
system, the scenario is different: For boundary rates for which a
stationary distribution exists condensation never occurs. Only in
the non-stationary case of very strong injection a dynamical
condensation phenomenon occurs: The boundary sites can become
supercritical, in which case the occupation number of these sites
grows indefinitely \cite{levine2005openzrp}. This gives rise to a
non-stationary condensate-like structure, but, in further contrast
to the usual stationary bulk condensate, this can happen for any
choice of bounded hopping rates $u(n)$ and, moreover, the phenomenon
is strictly limited to the boundary sites.

This brief outline of the properties of the ZRP concerns typical
behaviour of the stochastic dynamics. It is the purpose of this work
to study on microscopic scale the behaviour of the ZRP in a regime
of strongly atypical behaviour of the particle current. This is of
interest for several reasons. First, it serves not only as a test
for macroscopic fluctuation theory (MFT) \cite{Bert15}, but also
probes its validity in the regime of strong asymmetry where MFT is
not rigorous. Second, we will be able to compute effective
microscopic interactions that make atypical behaviour typical.
Third, it will transpire that unexpected and novel condensation
patterns can occur even when the unconditioned dynamics do not
exhibit condensation.

Current large deviations of finite duration have been investigated
for the ZRP in the context of the breakdown of the Gallavotti-Cohen
symmetry for the current distribution in a ZRP with open boundary
conditions \cite{HarrisEtal2006GallavottiCohenBreakdown,Rako08}. It
turns out that the failure of the Gallavotti-Cohen symmetry
argument, which is based on a very general time-reversal property of
stochastic dynamics \cite{Gall95,Lebo99}, can be related to the
formation of ``instantaneous condensates''
\cite{HarrisEtal2006GallavottiCohenBreakdown}. These condensates
were investigated recently in terms of Doob's $h$-transform for the
ZRP with a single site \cite{HarrisEtal2013DynamicsAtypicalZRP}. It
was shown that in some parameter regimes, Doob's $h$-transform fails
to represent the effective dynamics that make the large deviation
typical.

The microscopic large-deviation properties of the ZRP in the regime
of atypical currents that persist for a very long time have not yet
been explored. In this paper we address this problem. Moreover, we
do not limit ourselves a single site; rather, we consider the full
problem of open boundaries with any number of sites. It turns out
that Doob's $h$-transform can be computed exactly from a product
ansatz for the lowest eigenvector. Thus we are able to calculate the
effective interactions that make the current large deviations
typical. This turns out to be a ZRP with space-dependent hopping
bias. Interestingly, the effective process satisfies detailed
balance if and only of one conditions on vanishing macroscopic
current. These results are outside the scope of MFT.

Somewhat surprisingly the exact results show that {\it bulk}
condensation in the conditioned open system may occur, as is the
case for periodic boundary conditions under typical dynamics.
However, in contrast to typical behaviour (and to naive
expectation), the results suggest that a whole lattice segment may
become supercritical rather than just a single site. The segment
location and length are fixed by the current on which one conditions
rather than randomly fluctuating as one has for the condensate
position in the periodic case with typical dynamics.

Our exact results are valid for any asymmetry, including the weakly
asymmetric case. This allows for a comparison with the predictions
of the macroscopic fluctuation theory which we apply to the ZRP with
weak asymmetry. The microscopic results show that the MFT results
remain intact in the limit of finite asymmetry, a limit in which the
validity of the macroscopic approach is a-priori questionable.
Moreover, we obtain the scaling factor for the asymmetry that cannot
be computed from the MFT.

The paper is organized as follows: In Sec.\ 2 we introduce the model
and define the conditioned dynamics. In Sec.\ 3 we derive the exact
microscopic results for the $h$-transform. This yields the effective
dynamics and the spatial condensation patterns of the conditioned
process. In Sec.\ 4 we follow the macroscopic approach for the
weakly asymmetric case to compute the optimal macroscopic profile
that realizes a current large deviation.

\section{Open ZRP conditioned on an atypical current}
\label{highcurrent}

\subsection{Definition of the model}
\label{ss:df}

\begin{figure}
\begin{center}
\includegraphics*[width=0.6\textwidth]{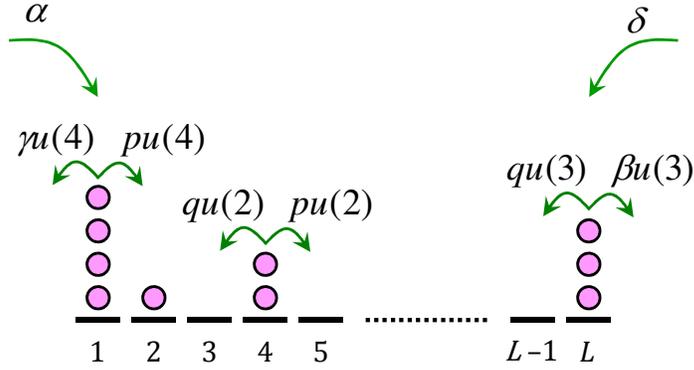}
\caption{Some possible particle hopping events in the ZRP with $L$ sites and
open boundary conditions.}
\label{f:ZRP}
\end{center}
\end{figure}

In the bulk of the lattice a particle from site $i$ hops to site
$i+1$ with rate $p u(n_i)$, and to site $i-1$ with rate $q u(n_i)$.
The parameters $p$ and $q$ determine the asymmetry of the process:
The hopping is symmetric when $p=q$ and asymmetric otherwise. For
the function $u(n)$ we have $u(0) = 0$, as particles cannot leave an
empty site, otherwise $u(n) >0$ is arbitrary. By definition the
$i$th bond is between sites $i$ and $i+1$ for $1 \leq i \leq L-1$.
Bond 0 ($L$) represents the link between the lattice and a left
(right) boundary reservoir that is not modeled explicitly. At the
left boundary, particles enter with rate $\alpha$ from the left
reservoir and exit with rate $\gamma u(n_1)$. Similarly, at the
right boundary particles enter with rate $\delta$ from the right
boundary reservoir and exit with rate $\beta u(n_L)$ (Fig.
\ref{f:ZRP}).

%\begin{figure}
%\centering
%\includegraphics[width=0.6\textwidth]{L200E02C05.eps}\vspace*{5mm}
%\includegraphics[width=0.6\textwidth]{L200E2C1.eps}\vspace*{5mm}
%\includegraphics[width=0.6\textwidth]{L200E2C05.eps} \vspace*{5mm}
%\includegraphics[width=0.6\textwidth]{L200E40C03.eps}
%\caption{Interpolated density profiles for type II shock measures for lattice size $L$=200, shock position
%$m=10$ and different asymmetries and total densities: (a) $E=0.2$, $c=0.5$,  (b) $E=2$, $c=1$, (c)
%$E=2$, $c=0.5$,  (d) $E=40$, $c=0.3$.}
%\label{typeIIshockprofiles}
%\end{figure}

The dynamics of the zero-range process can be conveniently
represented using the quantum Hamiltonian
formalism~\cite{Schu01,Lloy96}.  In this approach one defines a
probability vector $|P\rangle = \sum_\bfn P_\bfn \ket{\bfn}$ where
$\ket{\bfn} = |n_1) \otimes \dots \otimes |n_L)$ is the canonical
basis vector of $(\mathbb{C}^\infty)^{\otimes L}$ associated with
the particle configuration $\bfn=(n_1,n_2,\ldots,n_L)$ and $P_\bfn$
the probability measure on the set of all such configurations. For a
single site in this tensor product the configuration with $n$
particles on that site is represented by the basis vector $|n)$
which has component 1 at position $n+1$ and zero elsewhere. By
definition $|P\rangle$ obeys the normalization condition $\langle
\bfS |P \rangle =1$ with the summation vector \bel{sumvec}
\bra{\bfS} = \sum_\bfn  \bra{\bfn} \ee and the orthogonality
condition $\langle \bfn | \bfn' \rangle = \delta_{\bfn,\bfn'}$.
Within this formalism the Markovian time evolution of the ZRP is
represented by the Master equation
\begin{equation}
\label{ME}
\frac{\rmd}{\rmd t}|P(t)\rangle = - \hat{H} |P(t)\rangle
\end{equation}
with
\begin{eqnarray}
\hat{H} & = & \hat{h}_0 + \sum_{k=1}^{L-1} \hat{h}_k + \hat{h}_L \nonumber \\
    & = & - \biggl\{ \sum_{k=1}^{L-1} \left[  p(\hat{a}_k^- \hat{a}^+_{k+1} -\hat{d}_k) + q (\hat{a}_k^+ \hat{a}_{k+1}^- - \hat{d}_{k+1})\right] \nonumber \\
    &    & + \alpha(\hat{a}_1^+-1) + \gamma(\hat{a}_1^- - \hat{d}_1)  +\delta (\hat{a}_L^+-1)+\beta(\hat{a}^-_L-\hat{d}_L) \biggr\} \label{e:H}
\end{eqnarray}
where $\hat{a}^+$ and $\hat{a}^-$ are infinite-dimensional particle
creation and annihilation matrices
\begin{equation}
\fl \hat{a}^+=\left(\begin{array}{ccccc}
0 & 0 & 0 & 0 & \ldots \\
1 & 0 & 0 & 0 & \ldots \\
0 & 1 & 0 & 0 & \ldots \\
0 & 0 & 1 & 0 & \ldots \\
\ldots & \ldots & \ldots & \ldots & \ldots
\end{array}
\right),
\qquad
\hat{a}^-=\left(\begin{array}{ccccc}
0 & u(1) & 0 & 0 & \ldots \\
0 & 0 & u(2) & 0 & \ldots \\
0 & 0 & 0 & u(3) & \ldots \\
0 & 0 & 0 & 0 & \ldots \\
\ldots & \ldots & \ldots & \ldots & \ldots
\end{array} \label{e:aa+}
\right)
\end{equation}
and $\hat{d}= \sum u(r) |r)(r|$ is a diagonal matrix with the
$(r,s)$th element given by $u(r) \delta_{r,s}$. The subscript $k$ in
$\hat{a}^\pm_k$ and $\hat{d}_k$ indicates that the respective matrix
acts non-trivially on site $k$ of the lattice and as unit operator
on all other sites. We also introduce the local particle number
operator $\hat{n}_k$ with the diagonal particle number matrix
$\hat{n}  = \sum r |r)(r|$ and note the useful identity
\bel{particletrafo} y^{\hat{n}} \hat{a}^\pm y^{-\hat{n}} = y^{\pm 1}
\hat{a}^\pm \ee for any non-zero complex number $y$. The global
particle number operator $\hat{N} = \sum_{k=1}^L \hat{n}_k$ commutes
with the bulk part of $\hat{H}$, expressing particle number
conservation of the bulk jump processes.

According to (\ref{ME}) the probability vector at time $t$ is given
by \be \ket{P(t)} = \exp{(-\hat{H}t)}\ket{P(t)}. \ee Using
$(s|\hat{a}^+ = (s|$ and $(s|\hat{a}^- = (s| \hat{d}$ for the
one-site summation vector $(s| = \sum_n (n|$ one verifies that the
global summation vector \eref{sumvec} is a left eigenvector of
$\hat{H}$ with zero eigenvalue. This expresses conservation of
probability through the relation $ 0 = d/(dt) \inprod{\bfS}{P(t)} =
-\bra{\bfS}\hat{H}\ket{P(t)} $. A stationary distribution, denoted
$\ket{P^\ast}$, is a right eigenvector of $\hat{H}$ with eigenvalue
zero.

In~\cite{levine2005openzrp} it is shown that the stationary
distribution of the ZRP is given by a product measure
\begin{equation}
|P^*\rangle = |P^*_1) \otimes |P^*_2) \otimes \ldots \otimes |P^*_L) \label{e:prod}
\end{equation}
where the marginal distribution $|P^*_k)$ is the single-site probability vector with components
\begin{equation}
P^*_k := \Prob{n_k=n} = \frac{z_k^n}{Z_k}\prod_{i=1}^n u(i)^{-1} \label{e:prob}.
\end{equation}
The empty product is defined to be equal to 1 and
$Z_k$ is the local partition function
\begin{equation}
Z_k \equiv Z(z_k)=\sum_{n=0}^{\infty} z_k^n \prod_{i=1}^{n} u(i)^{-1}. \label{e:gc}
\end{equation}
The fugacities $z_k$ in the steady state are given by
\begin{equation}
z_k=\frac{[(\alpha+\delta)(p-q)-\alpha\beta+\gamma\delta]\left(\frac{p}{q}\right)^{k-1}
 - \gamma\delta + \alpha\beta \left(\frac{p}{q}\right)^{L-1}}
{\gamma(p-q-\beta) + \beta(p-q+\gamma)\left(\frac{p}{q}\right)^{L-1}}.
\label{e:zss2}
\end{equation}
The mean density at site $k$ is related to the fugacity by (see
(\ref{e:prob})--(\ref{e:gc}))
\begin{equation}\label{eq:ZRPdensity}
\rho_k = \langle n_k \rangle = \frac{d \log Z_k}{d \log z_k}
\end{equation}
The absence of detailed balance is reflected in a steady-state
current
\begin{equation}
j^\ast=(p-q)\frac{- \gamma\delta + \alpha\beta \left(\frac{p}{q}\right)^{L-1}}
{\gamma(p-q-\beta) + \beta(p-q+\gamma)\left(\frac{p}{q}\right)^{L-1}}.
\label{e:css}
\end{equation}

Notice that the existence of a steady state is determined by the
radius of convergence of $Z_k$.  If the form of the transition rates
results in some $z_k$ outside this radius of convergence, then a
stationary distribution does not exist. One expects relaxation to
the local marginal distribution wherever the radius of convergence
is finite, and a growing condensate on the other sites
\cite{levine2005openzrp}.

\subsection{Grandcanonical conditioning}

By ergodicity, the stationary current $j^\ast$ is the long-time mean
of the time-integrated current $J_k(t)$ across a bond $k,k+1$, i.e.
of the number of positive particle jumps from site $k$ to $k+1$
minus the number of negative particle jumps from site $k+1$ to $k$
up to time $t$. Because of bulk particle number conservation the
stationary current does not depend on $k$. It is of great interest
to consider not only the mean $j^\ast$, but also the fluctuations of
the time-averaged current $j_k(t) = J_k(t)/t$, in particular its
large deviations. The distribution $g_k(j,t) := \Prob{j_k(t) = j}$
has the large deviation property $g_k(j,t) \propto \rme^{-F_k(j)t}$
\cite{Lebo99}. For particle systems with finite state space the
large deviation function does not depend on the site $k$ and
satisfies the Gallavotti-Cohen symmetry \cite{Gall95,Lebo99} and
this is widely believed to be generic. However, it has been pointed
out that neither statement remains valid for the ZRP with open
boundaries if one considers sufficiently large atypical currents. In
order to get deeper insight into this breakdown we study here the
ZRP conditioned to produce a strong atypical current. However,
instead on enforcing a particular (integer) value of the
time-integrated current, we consider a ``grandcanonical''
conditioning defined in terms of a Legrende transform with conjugate
parameter $s$. This corresponds to an ensemble of ZRP-histories
where the current is allowed to fluctuate around some atypical mean.
We refer to this ensemble as the $s$-ensemble.

Grandcanonically conditioned Markov processes may be studied in the
spirit of Doob's $h$-transform \cite{Mill61,Doob}, as was done for
the ASEP conditioned on very large currents in
\cite{Simo09,PopkovEtal2010AsepEnhancedFlux,Popk11}, see also
\cite{Jack10,Chet13} for recent discussions. This microscopic
approach also provides a construction to compute effective
interactions for which the atypical behaviour becomes typical, see
\cite{Simo09,PopkovEtal2010AsepEnhancedFlux} for applications to the
ASEP. Besides the conceptual insight gained in this way into extreme
behaviour, this is potentially interesting from a practical
perspective: By adjusting interactions between particles, one
obtains a means to make desirable, but normally rare, events
frequent.

Following \cite{Derr98,Harr07} this construction is done by first
defining a weighted generator $\hat{H}(s)$ where the operators that
correspond to a jump to the right (left) are multiplied by a factor
$\rme^{s}$ ($\rme^{-s}$). Then one considers the lowest left
eigenvector of $\hat{H}(s)$, i.e., the eigenvector $\bra{0}$ to the
lowest eigenvalue of $\hat{H}(s)$, which we denote by
$\epsilon_0(s)$. We also introduce the diagonal matrix $ \hat{D}(s)$
which has the components $ D_\bfn(s)$ of the lowest eigenvector on
its diagonal. Since all off-diagonal elements of $\hat{H}(s)$ are
non-positive we appeal to Perron-Frobenius theory and argue that all
components of
%$\bra{ \hat{D}(s)}$
$\bra{0}$ can be chosen to be real and non-zero. Thus $ \hat{D}$ is
invertible and we have \bel{Delta} \bra{0} = \bra{\bfS} \hat{D}(s),
\quad \bra{0}  \hat{D}^{-1}(s) = \bra{\bfS}. \ee This allows us to
introduce the grandcanonical Doob's $h$-transform \bel{htrafo}
\hat{G}(s)  =  \hat{D}(s) \hat{H}(s)    \hat{D}^{-1}(s) -
\epsilon_0(s). \ee Notice that by construction $\hat{G}(s)$ has
non-positive off-diagonal elements and, moreover, by \eref{Delta},
$\bra{\bfS} \hat{G}(s) = 0$. Hence $\hat{G}(s)$ is the generator of
some Markov process, which we shall refer to as the effective
generator, with jump rates $\tilde{w}_{\bfn', \bfn}(s) =  w_{\bfn',
\bfn} D_{\bfn'}(s)  D^{-1}_{\bfn}(s)$. We denote by $P^\ast_\bfn(s)$
the invariant measure of the effective process $ \hat{G}(s)$. The
steady state of this Markov process provides the steady state of the
grandcanonically biased $s$ model. It is easy to prove that
$P^\ast_\bfn(s) = D_\bfn(s) \Gamma_\bfn(s)$ where $\Gamma_\bfn(s)$
are the components of the lowest right eigenvector of the weighted
generator $\hat{H}(s)$. In order to avoid heavy notation we suppress
in the following the dependence on $s$ if there is no danger of
confusion.

\subsection{Local conditioning in the ZRP}

Under the condition that the mean integrated current across some
fixed bond $(k,k+1)$ fluctuates around a certain value parameterized
by $s$ we obtain, following Ref. \cite{Harr07}, the weighted
generator \bel{fullmatrix} \hat{H}^{(k)}(s) = \sum_{l=0}^{k-1}
\hat{h}_l + \hat{h}_k(s) + \sum_{l=k+1}^{L} \hat{h}_l . \ee Here
\bea
\hat{h}_0(s) & = & -\left[\alpha(\rme^{s} \hat{a}_1^+  - 1) + \gamma(\rme^{-s} \hat{a}_1^-  - \hat{d}_1)\right] \nonumber\\
\hat{h}_k(s) & = & -\left[  p(\rme^{s} \hat{a}_k^- \hat{a}^+_{k+1}
-\hat{d}_k) + q (\rme^{-s} \hat{a}_k^+ \hat{a}_{k+1}^- -
\hat{d}_{k+1})\right],
\quad 1\leq k \leq L-1 \nonumber \\
\hat{h}_L(s) & = & -\left[ \delta
(\rme^{-s}\hat{a}_L^+-1)+\beta(\rme^{s}\hat{a}^-_L-\hat{d}_L)
\right]  . \eea We refer to this setting as {\it local
conditioning}. Notice that $\hat{h}_k(0) = \hat{h}_k$. We denote the
lowest left eigenvector of $\hat{H}^{(k)}$ by $\bra{Y_k} =
\bra{\bfS}  \hat{D}_k$. Specifically for $k=0$ we drop the subscript
$0$ i.e., we write $\bra{Y_0} = \bra{Y} = \bra{\bfS}  \hat{D}$

Define the partial number operator $\hat{N}_k = \sum_{i=1}^k
\hat{n}_i$. From \eref{particletrafo} one concludes
$\hat{H}^{(k)}(s) = \rme^{-s \hat{N}_k} \hat{H}^{(0)}(s) \rme^{s
\hat{N}_k}$. This implies for the left eigenvector \be \bra{Y_k}
\hat{H}^{(k)}(s) = \bra{Y_k} \epsilon_0 \ee where the lowest
eigenvalue is independent of $k$ and \be \bra{Y_k} = \bra{Y} \rme^{s
\hat{N}_k} = \bra{\bfS}  \hat{D}_k \ee with $ \hat{D}_k =  \hat{D}
\rme^{s \hat{N}_k}$. This yields effective dynamics \be
\hat{G}^{(k)}(s) =  \hat{D}_k \hat{H}^{(k)}(s)   \hat{D}_k^{-1} -
\epsilon_0 =  \hat{D} \hat{H}^{(0)}(s)   \hat{D}^{-1} - \epsilon_0 =
\hat{G}^{(0)}(s). \ee We conclude that the effective dynamics does
not depend on $k$ and we can focus on conditioning on a current
across bond $0$, i.e. between the left reservoir and site 1 of the
lattice.

\section{Microscopic density profiles and condensation}

\subsection{Left eigenvector}

The left eigenvector of  $\hat{H}^{(0)}(s)$ was computed in
\cite{HarrisEtal2006GallavottiCohenBreakdown}. For
self-containedness we repeat here the essential steps, which are
based on a product ansatz $\bra{Y} = (y_1| \otimes (y_2| \otimes
\ldots \otimes (y_L|  = \bra{\bfS}  \hat{D}$ with the diagonal
matrix $ \hat{D}=y_1^n \otimes \ldots y_L^n $ where $n$ are the
diagonal one-site number operators. Then \bea
\fl - \bra{Y} \hat{H}^{(0)}(s) & = & \bra{Y} \left\{ \alpha (y_1\rme^s-1) + \gamma(\rme^{-s}-y_1) \hat{d}_1y_1^{-1} + \right. \nonumber\\
&   & \sum_{k=1}^{L-1}\left[ p (y_{k+1}-y_k) \hat{d}_ky_k^{-1} + q (y_k-y_{k+1}) \hat{d}_{k+1} y_{k+1}^{-1} \right]  + \\
&   & \left. \delta (y_L-1) + \beta (1-y_L) \hat{d}_Ly_L^{-1} \right\} \nonumber\\
& = & \bra{Y} \left\{ \alpha (y_1\rme^s-1) + \left[\gamma(\rme^{-s}-y_1) + p(y_2-y_1)\right]\hat{d}_1y_1^{-1} + \right. \nonumber\\
&   & \sum_{k=2}^{L-1}\left[ p (y_{k+1}-y_k)  + q (y_{k-1}-y_k)  \right]  \hat{d}_ky_k^{-1} + \\
&   & \left. \delta (y_L-1) + \left[ q (y_{L-1}-y_L) + \beta
(1-y_L)\right] \hat{d}_Ly_L^{-1} \right\}. \nonumber \eea One sees
that $\bra{Y}$ is a left eigenvector if the following equations are
satisfied: \bea \label{bulkrecursion}
0 & = &  p (y_{k+1}-y_k)  + q (y_{k-1}-y_k) \\
\label{leftrecursion}
0 & = & \gamma(\rme^{-s}-y_1) + p(y_2-y_1)\\
\label{rightrecursion} 0 & = & q (y_{L-1}-y_L) + \beta (1-y_L). \eea
We define the hopping asymmetry $a = p/q$. The ansatz \bel{ansatzy}
y_k = A + B a^{L+1-k} \ee yields $y_{k-1}-y_k = B (a-1) a^{L+1-k}$
and therefore solves \eref{bulkrecursion}.

The left boundary equation \eref{leftrecursion} yields \be \rme^{-s}
- A - Ba^{L} + p B a^{L} (a^{-1}-1)/\gamma=0 \ee which gives \be A =
\rme^{-s} - (p-q+\gamma)\frac{B a^L}{\gamma}. \ee On the other hand,
the right boundary equation \eref{rightrecursion} yields \be B (p-q)
a + \beta(1-A-B a) = 0 \ee Plugging this into $A$ leads to
\bel{constB} B =
\frac{\beta\gamma(\rme^{-s}-1)a^{-1}}{\gamma(p-q-\beta)+\beta(p-q+\gamma)a^{L-1}}
\ee and \bel{constA} \fl A =
\frac{\gamma\rme^{-s}(p-q-\beta)+\beta(p-q+\gamma)a^{L-1}}{\gamma(p-q-\beta)+\beta(p-q+\gamma)a^{L-1}}
= 1+
\frac{\gamma(\rme^{-s}-1)(p-q-\beta)}{\gamma(p-q-\beta)+\beta(p-q+\gamma)a^{L-1}}.
\ee

For the lowest eigenvalue we get
\bea
\epsilon_0 & = & - ( \alpha (y_1 \rme^s - 1) + \delta (y_L-1)) \nonumber\\
& = & \frac{\alpha}{\gamma} p \rme^s (y_1-y_2) - \frac{\delta}{\beta} q (y_{L-1}-y_L)\\
& = & (p-q)(\frac{\alpha}{\gamma} a^{-1} \rme^s - \frac{\delta}{\beta} a^{-L})Ba^{L+1} \nonumber
\eea
from which one finds
\be
\epsilon_0 = (p-q)(\rme^{-s} - 1)
\frac{\alpha\beta a^{L-1} \rme^s - \gamma\delta}{\gamma(p-q-\beta)+\beta(p-q+\gamma)a^{L-1}}.
\ee

\subsection{Right eigenvector}

Following \cite{HarrisEtal2006GallavottiCohenBreakdown} we make a
product ansatz also for the right eigenvector: $\ket{X} = |x_1)
\otimes |x_2) \otimes \ldots \otimes |x_L) $ where $|x_k)$ is the
unnormalized vector with components
\begin{equation}
(Q^*_k)_n = x_k^n \prod_{i=1}^n u(i)^{-1} .
\end{equation}
This yields \bea
- \hat{h}_k \ket{X} & = & (p x_k - q x_{k+1}) \left( \frac{\hat{d}_k}{x_k} - \frac{\hat{d}_{k+1}}{x_{k+1}} \right)\ket{X}  \\
- \hat{h}_0(s) \ket{X} & = & (\alpha \rme^s - \gamma x_1) \left(\rme^{-s} - \frac{\hat{d}_1}{x_1} \right) \ket{X} \\
- \hat{h}_L \ket{X}  & = & (\beta x_L - \delta) \left( 1 -
\frac{\hat{d}_{L}}{x_{L}} \right) \ket{X} \eea
One sees that
$\ket{X}$ is a right eigenvector if the following equations are
satisfied: \bea \label{bulkrecursion2}
b & = &  p x_k  - q x_{k+1} \\
\label{leftrecursion2}
b & = & \alpha\rme^{s}-\gamma x_1 \\
\label{rightrecursion2} b & = & \beta x_L - \delta
\eea
with some constant $b$. The ansatz
\bel{ansatzx}
x_k = C + D a^{k}
\ee
yields $p x_k - q x_{k+1} = C (p-q)$ and therefore solves \eref{bulkrecursion} with
$b= C (p-q)$. Using the boundary recursions one then obtains
\bel{constC}
C = \frac{\alpha\beta \rme^s a^{L-1} - \gamma\delta}{\beta(p-q+\gamma)
a^{L-1} + \gamma(p-q-\beta)}
\ee
and
\bel{constD}
D = a^{-1} \frac{\alpha
\rme^s (p-q-\beta) + \delta (p-q+\gamma)}{\beta(p-q+\gamma) a^{L-1}
+ \gamma(p-q-\beta)} .
\ee
Notice that in terms of the local
fugacities $x_k$ one has
\bel{energyexpression}
\epsilon_0 = -\alpha
+ \gamma \rme^{-s} x_1 -\delta + \beta x_L = C (1-\rme^{-s}).
\ee

\subsection{Effective dynamics}

The transformation \eref{particletrafo} yields for the bulk hopping
terms \bea -  \hat{D} \hat{h}_k  \hat{D}^{-1} & = & p
\left(\frac{y_{k+1}}{y_k} \hat{a}_k^- \hat{a}_{k+1}^+ -
\hat{d}_k\right)
+ q \left(\frac{y_k}{y_{k+1}} \hat{a}_k^+ \hat{a}_{k+1}^- - \hat{d}_{k+1}\right) \nonumber\\
 & = & p \frac{y_{k+1}}{y_k} \left( \hat{a}_k^- \hat{a}_{k+1}^+ - \hat{d}_k\right)
+ q \frac{y_k}{y_{k+1}} \left(\hat{a}_k^+ \hat{a}_{k+1}^- - \hat{d}_{k+1}\right) \\
& & + (p-q) \left(1-\frac{A}{y_{k+1}}\right)\hat{d}_{k+1} - (p-q)
\left(1-\frac{A}{y_{k+1}}\right)\hat{d}_k \nonumber \eea and for the
boundaries \bea
-  \hat{D} \hat{h}_0  \hat{D}^{-1} & = & \alpha(y_1 \rme^{s} \hat{a}_1^+  - 1) + \gamma( y_1^{-1} \rme^{-s} \hat{a}_1^-  - \hat{d}_1) \nonumber\\
% & = & \alpha y_1 \rme^{s} ( \hat{a}_1^+ - 1) + \gamma y_1^{-1} \rme^{-s} ( \hat{a}_1^-  - \hat{d}_1)
% +  \alpha (y_1 \rme^{s} - 1) +  \gamma y_1^{-1} ( \rme^{-s}  - y_1) \hat{d}_1 \\
 & = & \alpha y_1 \rme^{s} ( \hat{a}_1^+ - 1) + \gamma y_1^{-1} \rme^{-s} ( \hat{a}_1^-  - \hat{d}_1) \\
 & &
 +  \alpha (y_1 \rme^{s} - 1) +  (p-q) y_1^{-1} ( y_1 - A) \hat{d}_1 \nonumber
\eea and \bea
-  \hat{D} \hat{h}_L  \hat{D}^{-1} & = & \delta (y_L \hat{a}_L^+ - 1)+\beta(y_L^{-1} \hat{a}^-_L -\hat{d}_L) \nonumber\\
%& = & \delta y_L (\hat{a}_L^+ - 1)+\beta y_L^{-1} (\hat{a}^-_L -\hat{d}_L)
%+ \delta (y_L  - 1)+\beta y_L^{-1} (1-y_L) \hat{d}_L \\
& = & \delta y_L (\hat{a}_L^+ - 1)+\beta y_L^{-1} (\hat{a}^-_L -\hat{d}_L) \\
& & + \delta (y_L  - 1)+ (p-q) y_L^{-1} (y_L-A) \hat{d}_L \nonumber
\eea Therefore the effective dynamics is given by \bea
\label{effdyn}
\hat{G}^{(0)}(s) & = & \hat{D} \hat{G}^{(0)}  \hat{D}^{-1} - \epsilon_0 \nonumber \\
& = & - \sum_{k=1}^{L-1} \left[ p \frac{y_{k+1}}{y_k} \left(
\hat{a}_k^- \hat{a}_{k+1}^+ - \hat{d}_k\right)
+ q \frac{y_k}{y_{k+1}} \left(\hat{a}_k^+ \hat{a}_{k+1}^- - \hat{d}_{k+1}\right) \right] \\
& & - \left[ \alpha y_1 \rme^{s} ( \hat{a}_1^+ - 1) + \gamma y_1^{-1} \rme^{-s} ( \hat{a}_1^-  - \hat{d}_1) \right] \nonumber\\
& & - \left[ \delta y_L (\hat{a}_L^+ - 1)+\beta y_L^{-1}
(\hat{a}^-_L -\hat{d}_L) \right] \nonumber \eea It is intriguing
that this is a driven ZRP with a spatially varying driving field
$E_k(s) = \log{a} + 2 \log{ y_{k+1}(s)/y_k(s)}$. This
space-dependence will be present even in the case of non-interacting
particles with $u(n)= wn$. Therefore, conditioning on an atypical
local current can be realized by an effective process with a
space-dependent driving field.

The stationary distribution of the effective dynamics is a product
state
\bel{effstationary} \ket{P^*(s)} =  |P^*_1(s) ) \otimes
|P^*_2(s) ) \otimes \ldots \otimes |P^*_L(s)) \ee where $|P^*_k(s))$
is the probability vector with components
\begin{equation}
(P^*_k(s))_n = \frac{z_k^n}{Z_k}\prod_{i=1}^n u(i)^{-1} .
\end{equation}
Here \bel{zk} z_k = x_k y_k \ee is the local fugacity given by
\eref{ansatzy} together with \eref{constB}, \eref{constA} and
\eref{ansatzx} together with \eref{constC} \eref{constD}. The
normalization $Z_k$ is the local analogue of the grand-canonical
partition function, given by (\ref{e:gc}) as before. The density at
site $k$ is related to the fugacity as in the unconditioned ZRP by
\eqn (\ref{eq:ZRPdensity}). For example, for non-interacting
particles $u(n) = n$, and therefore the density is simply $\rho_k =
z_k$. It is remarkable that one obtains a non-trivial density
profile even for non-interacting particles.

The stationary current is \bel{currenteffective} \fl j^\ast(s) = p
x_k y_{k+1} - q y_k x_{k+1} = \alpha y_1 \rme^{s} - \gamma x_1
\rme^{-s} = \beta x_L - \delta y_L = (p-q) (AC-BDa^{L+1}) . \ee
Plugging in the expressions yields \bel{currenteffective2} j^\ast
(s) = (p-q) \frac{\alpha\beta\rme^s a^{L-1} - \gamma\delta\rme^{-s}}
{\beta(p-q+\gamma) a^{L-1} + \gamma(p-q-\beta)} . \ee

\subsection{Examples}

We now explore in more depth the results of the previous subsections
for some specific choices of the model parameters.

\subsubsection{Barrier-free reservoirs}

To somewhat reduce the parameter space of the model, it is natural
to consider
\begin{eqnarray}
\label{nat1} \beta &=& p \\
\label{nat2} \gamma &=& q,
\end{eqnarray}
i.e., the hopping rates out of the first and last site are the same
as the bulk hopping rates. We further simplify the notation by
reexpressing the parameters $\alpha$ and $\delta$ in terms of
reservoir fugacities:
\begin{eqnarray}
\label{nat3} \alpha &=& z_l\, \gamma\, p / q = z_l\,p \\
\label{nat4} \delta &=& z_r\, \beta\, q / p = z_r\, q
\end{eqnarray}
where $z_l$ is the fugacity of the left reservoir and $z_r$ is that
of the right reservoir. The interpretation is that in the original
(unconditioned) process, particles jump with the same rate between
reservoir and boundary sites as inside the bulk of the chain. We
therefore refer to this choice of rates as ``barrier-free
reservoirs''.

A few further notations will prove useful in what follows. First, we
express the hopping asymmetry $a = p/q$ via
\begin{equation}\label{eq:nutilde}
\nutilde = \frac{L+1}{2} \log a.
\end{equation}
It is further convenient to characterize the boundary reservoirs by
chemical potentials $z_i = \rme^{\mu_i}$, and to define parameters $
\Delta \mu$, $\bar{\mu}$ and $z_0$ as follows
\begin{eqnarray}\label{eq:deltamu}
\Delta \mu &\equiv& \mu_l - \mu_r = \log(z_l/z_r), \\
\bar{\mu} &\equiv& (\mu_l + \mu_r)/2, \label{eq:mubar}\\
z_0 &\equiv& \rme^{\bar{\mu}} = \sqrt{z_l z_r}. \label{eq:z0}
\end{eqnarray}
Finally, we define a relative position along the lattice
\begin{equation}\label{eq:rk}
\rk{k} = \frac{k}{L+1}.
\end{equation}

\subsubsection{Partially asymmetric ZRP with barrier-free reservoirs}

With the choice \eref{nat1}--\eref{nat4} the parameters $A,B,C,D$
take the simpler form \bea
A = \frac{a^{L+1}-\rme^{-s}}{a^{L+1}-1} &\qquad & B =   \frac{\rme^{-s}-1}{a^{L+1}-1} \\
C = \frac{z_{l} \rme^{s} a^{L+1}-z_r}{a^{L+1}-1} & & D = \frac{z_r -
z_{l} \rme^{s} }{a^{L+1}-1}. \eea This leads to \be \fl y_k =  1 -
(1- \rme^{-s})\frac{a^{L+1-k}-1}{a^{L+1}-1}  =
\frac{\rme^{\nutilde(1-\rk{k})}\sinh[\nutilde \rk{k}] +
\rme^{-s-\nutilde
\rk{k}}\sinh[\nutilde(1-\rk{k})]}{\sinh[\nutilde]}. \ee The
effective hopping rates can now be read off \eqn (\ref{effdyn}).
They corresponds to a ZRP in which the bulk bias to jump to the left
and right is site dependent: a particle hops from site $k$ to $k+1$
with rate $u(n_k) p_k$ and from site $k$ to $k-1$ with rate $u(n_k)
q_k$ where
\begin{equation}\label{eq:effectivep}
p_k = p \frac{y_{k+1}}{y_k} = \sqrt{pq}\,\frac{\rme^{s+\nutilde}\sinh[\nutilde \rk{k+1}] +
\sinh[\nutilde(1-\rk{k+1})]}{\rme^{s+\nutilde}\sinh[\nutilde \rk{k}] +
\sinh[\nutilde(1-\rk{k})]}
\end{equation}
and
\begin{equation}\label{eq:effectiveq}
q_k = q \frac{y_{k-1}}{y_{k}} = \sqrt{pq}\,\frac{\rme^{s+\nutilde}\sinh[\nutilde \rk{k-1}] +
\sinh[\nutilde(1-\rk{k-1})]}{\rme^{s+\nutilde}\sinh[\nutilde \rk{k}] +
\sinh[\nutilde(1-\rk{k})]}.
\end{equation}
These rates correspond to a space-dependent local field
\begin{equation}\label{eq:effectiveField}
E_k = \log \frac{p_k}{q_{k+1}} =  2 \log \frac{\rme^{s+\nutilde}\sinh[\nutilde \rk{k+1}] +
\sinh[\nutilde(1-\rk{k+1})]}{\rme^{s+\nutilde}\sinh[\nutilde \rk{k}] +
\sinh[\nutilde(1-\rk{k})]}.
\end{equation}
It is interesting to note that the effective bulk hopping rates, and
thus the effective field, are independent of the reservoir
fugacities.

At the left boundary (site 1), according to (\ref{effdyn}),
particles are effectively injected with rate $z_l\,p_0$ and removed
with rate $q_1$, where $p_0, q_1$ are given in (\ref{eq:effectivep})
and (\ref{eq:effectiveq}). Similarly at the right boundary (site
$L$) the effective injection and removal rates are $z_r\,q_{L+1}$
and $p_L$, respectively. Thus, the effective boundary dynamics
remains of the barrier-free form with the same reservoir fugacities,
and only the effective space-dependent hopping rates depend on the
current-conditioning.

\begin{figure}
\begin{center}
\includegraphics*[width=0.99\textwidth]{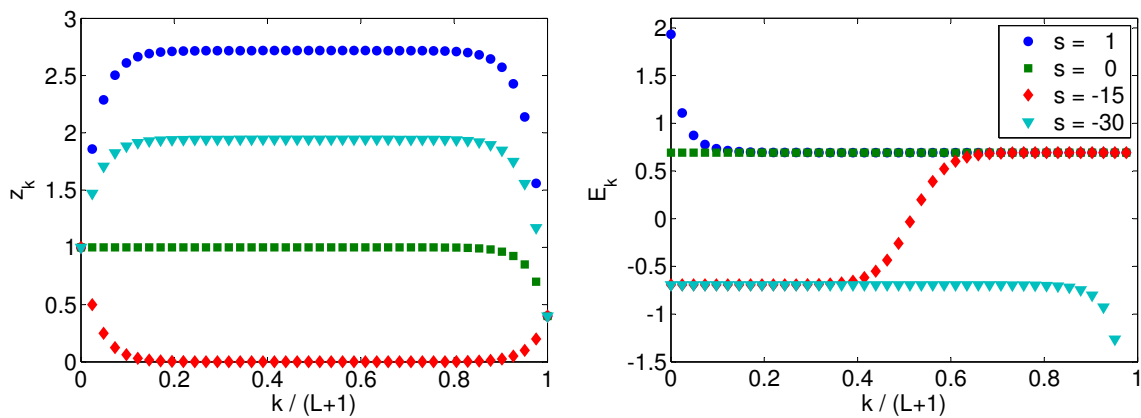}
\caption{The fugacity profile $z_k$ (left) and the effective field $E_k$
        (right) for an asymmetric ZRP with $p = 1$, $q=1/2$, $z_l = 1$,
        $z_r = 0.4$, and $L=40$. Several values of $s$ are plotted including
        the typical profile of the unconditioned process ($s=0$). The
        fugacity profile and the effective field are both flat, except for boundary
        layers of size $O(1)$ sites.}
\label{f:StrongAsym}
\end{center}
\end{figure}

\begin{figure}
\begin{center}
\includegraphics*[width=0.99\textwidth]{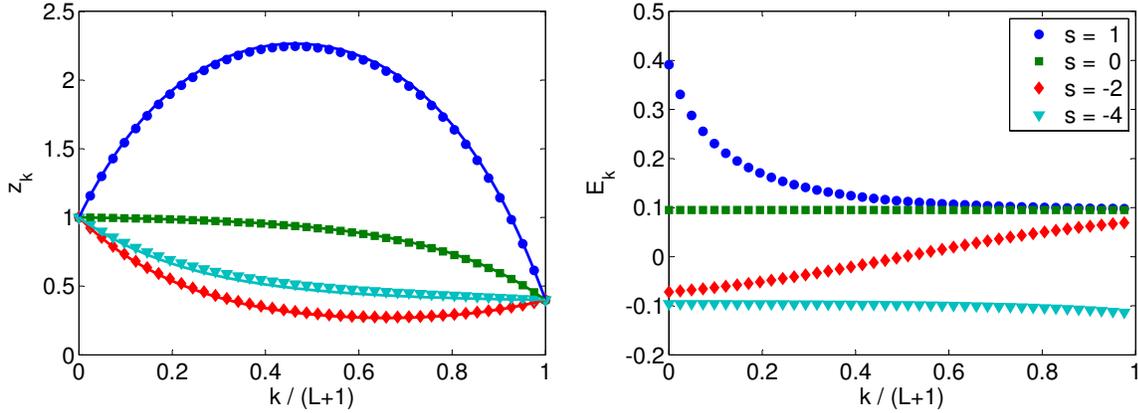}
\caption{The fugacity profile $z_k$ (left) and the effective field $E_k$
        (right) for a weakly-asymmetric ZRP with $p = 1/2 + 2/L$, $q=1/2$, $z_l = 1$,
        $z_r = 0.4$, and $L=40$. Several values of $s$ are plotted including
        the typical profile of the unconditioned process ($s=0$). The solid lines
        correspond to the macroscopic result, \eqns (\ref{eq:OptimalProfileJ})
        and (\ref{eq:QofS}).}
\label{f:WeakAsym}
\end{center}
\end{figure}

We proceed to examine the typical fugacity profile during the
atypical current event. To this end, we first calculate the right
eigenvector
\begin{equation}
\fl x_k = \frac{z_{l} \rme^{s} a^{L+1} + (z_r - z_{l}
\rme^{s})a^k -z_r}{a^{L+1}-1}= \frac{z_l \rme^{s+\nutilde
\rk{k}}\sinh[\nutilde(1-\rk{k})]+z_r \rme^{-\nutilde(1-\rk{k})}\sinh[\nutilde\rk{k}]}
{\sinh[\nutilde]}.
\end{equation}
Thus, the fugacity profile $z_k = x_k y_k$ is
\begin{equation}
\label{eq:OptimalProfileJmicroscopic}
\fl z_k = z_0 \frac{\rme^{\frac{  \Delta \mu}{2}} \sinh^2[\nutilde(1-\rk{k})] +
\rme^{-\frac{  \Delta \mu} {2}} \sinh^2[\nutilde \rk{k}]
+ 2 \tilde{Q} \sinh[\nutilde \rk{k}]\sinh[\nutilde(1-\rk{k})]}{\sinh^2\tilde{\nu}},
\end{equation}
with
\begin{equation}\label{eq:Qtilde}
\tilde{Q} = \cosh\Bigl(s +\frac{  \Delta \mu}{2} + \nutilde\Bigr).
\end{equation}
The fugacity profile (\ref{eq:OptimalProfileJmicroscopic}) along
with the effective driving field are plotted in figure
\ref{f:StrongAsym} for strongly asymmetric dynamics, and in figure
\ref{f:WeakAsym} for weakly asymmetric dynamics.

The current is \bel{currbf} j^\ast (s) = (p-q) \frac{z_{l}\rme^s
a^{L+1} - z_{r}\rme^{-s}} { a^{L+1} -1} =
2\sqrt{pq}\,\frac{\sinh(\frac{\nutilde}{L+1})\sinh\bigl(\nutilde+s+
\frac{\Delta\mu}{2}\bigr)}{\sinh(\nutilde)}. \ee This current could
be generated by a ZRP with the same constant bulk hopping rates
$p,q$ as in the original (unconditioned) dynamics but with effective
reservoir fugacities $z_l^\mysubtext{eff} = z_{l} \rme^{ s}$ and
$z_r^{\mysubtext{eff}} = z_{r}\rme^{-s}$. However, as pointed out
above, in a finite system this is {\it not} how the conditioned
process is typically realized. There is a space-dependent driving
field and the effective fugacities $z_k$ are not those of an
effective left reservoir with boundary fugacity $z_{l} \rme^{s}$.

\subsubsection{Symmetric ZRP}\label{sec:symmetric}

For $p=q$ both $A$ and $B$ of \eqns (\ref{constB}) and
(\ref{constA}) diverge, but $y_k$ is well-defined. One gets \be y_k
= 1 + \frac{\gamma(\rme^{-s}-1)}{p(\beta+\gamma)+\beta\gamma(L-1)}
[p + \beta (L-k)] \ee and similarly \be x_k = \frac{\alpha \rme^s
[p+\beta(L-k)] +
\delta[p+\gamma(k-1)]}{p(\beta+\gamma)+\beta\gamma(L-1)}. \ee For
the current this yields \be j^\ast (s) =
 p \frac{\alpha\beta\rme^s  - \gamma\delta\rme^{-s}}
{p(\beta+\gamma)  + \beta \gamma(L-1)}. \ee

In the case of barrier-free reservoirs of $\beta=\gamma=p$, $\alpha
= z_l p$, and $\delta = z_r p$, there is a simplification for which
one finds \be y_k = \rme^{-s}(1-\rk{k}) +  \rk{k} \qquad
\mytext{and} \qquad x_k = z_l \rme^s(1-\rk{k}) + z_{r} \rk{k}. \ee
The fugacity profile is therefore quadratic,
\begin{equation}
z_k = z_0 \Bigl[\rme^{\frac{  \Delta \mu}{2}} (1-\rk{k})^2 + \rme^{-\frac{ \Delta \mu} {2}}
\rk{k}^2
+ 2 \cosh\Bigl(s + \frac{  \Delta \mu}{2}\Bigr) \rk{k} (1-\rk{k}) \Bigr],
\end{equation}
and the current reduces to \be j^\ast (s) =
 p \frac{z_l\rme^s  - z_r\rme^{-s}}{L+1}.
\ee The effective field (\ref{eq:effectiveField}) in this case is
\begin{equation}\label{eq:SymmetricField}
E_k = 2 \log \frac{\rme^{s/2}\, \rk{k+1} + \rme^{-s/2}\,(1-\rk{k+1})}
{\rme^{s/2}\, \rk{k} + \rme^{-s/2}\,(1-\rk{k})}.
\end{equation}

Specifically, for $z_l=z_r=:z_0$ the unconditioned ZRP is in
equilibrium. Nevertheless, the conditioned process has non-trivial
properties. There is a current $j^\ast (s) = 2 p z_0 \sinh(s)/(L+1)$
and the fugacity profile
\begin{equation}\label{eq:SymmetricEffectiveProfile}
z_k = z_0 \left[1 + 4 \rk{k}\left(1-\rk{k}\right)
\sinh^2\left(\frac{s}{2}\right) \right]
\end{equation}
depends on space in a non-linear (quadratic) fashion. The
conditioned process has a space-dependent drift $E_k =
\sinh(s/2)[\rme^{s/2}\,\rk{k}+\rme^{-s/2}\,(1-\rk{k})]^{-1}$ which
decays algebraically with distance from the boundaries.

\begin{figure}
\begin{center}
\includegraphics*[width=0.99\textwidth]{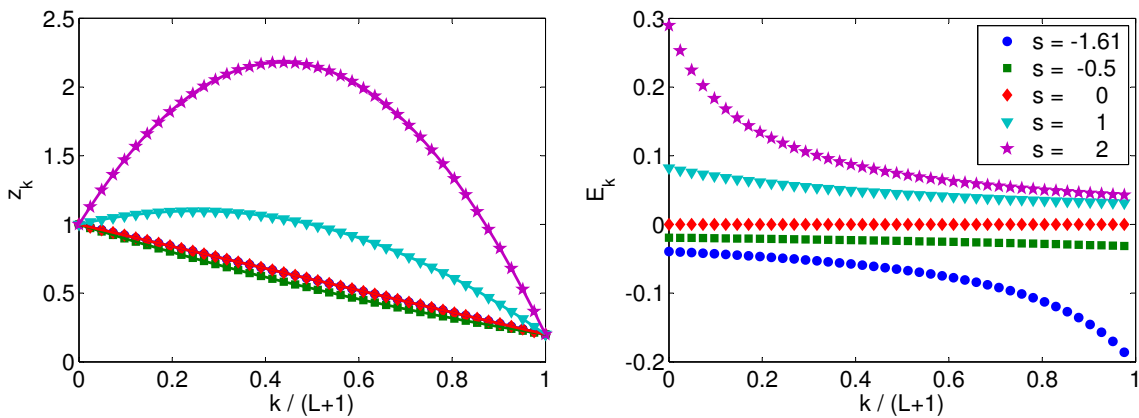}
\caption{The fugacity profile $z_k$ (left) and the effective field $E_k$
        (right) for a symmetric ZRP with $p = q = 1/2$, $z_l = 1$,
        $z_r = 0.2$, and $L=40$. Several values of $s$ are plotted including
        the typical profile of the unconditioned process ($s=0$) where the fugacity
        profile is linear. When $s = -\Delta \mu \approx -1.61$, the fugacity
        profile is linear again (and coincides with that of $s=0$), but the
        corresponding effective field is non-zero and space-dependent. Solid lines
        correspond to the macroscopic result, \eqn (\ref{eq:OptimalSymmetricS}).}
\label{f:sym}
\end{center}
\end{figure}

The fugacity profile and effective field for a symmetric ZRP are
presented in figure \ref{f:sym}.

\subsubsection{Totally asymmetric ZRP}

Consider $q=\gamma=\delta=0$ where particles can jump only to the right.
One has $A=1$ and $B=0$ (which is always true if $\gamma=0$). On the other hand,
$C=\alpha \rme^s /p$ and $D a^k = 0$ for $1 \leq k \leq L-1$ and $D a^L = \alpha \rme^s
(p-\beta)/(p\beta)$.
Therefore
\bea
x_k = z_k & = & \alpha \rme^s /p  \quad (1 \leq k \leq L-1) \\
x_L = z_L & = & \alpha \rme^s /\beta. \eea For the current one has
\be j^\ast (s) =  \alpha \rme^s . \ee As above one may parametrize
the boundary rates in terms of boundary fugacities $\alpha = z_{l}
p$ and $\beta = z_{r} p$. Then $z_k = z_L \rme^s$ for $1 \leq k \leq
L-1$ and $z_L = z_{l} \rme^s/z_{r}$. For the totally asymmetric
case, the effective dynamics is that of the original process, but
with an effective reservoir fugacity $z_{l}^{\mysubtext{eff}} =
z_{l} \rme^s$.

\subsection{Spatial condensation patterns}

At this point we recall that the construction presented above is
valid only for normalizable left and right eigenvectors, i.e., if
$\inprod{Y}{P_0} <\infty$ for any given initial distribution
$\ket{P_0}$, $\inprod{S}{X} <\infty$ and $Z_k <\infty$. In
particular, this is the case for non-interacting particles where
$z_k$ is the stationary local density. On the other hand, depending
on the choice of model parameters, the normalization condition can
be violated. In this case, a stationary distribution does not exist
and one expects the emergence of condensates
\cite{HarrisEtal2006GallavottiCohenBreakdown,Rako08,HarrisEtal2013DynamicsAtypicalZRP}.
It is intriguing that because of the space-dependence of the local
fugacities the existence of such condensates will generically depend
on the lattice site $k$ and condensates may form in the bulk of the
system.

To illustrate how conditioning on atypical currents may lead to
condensation we focus on the example of symmetric hopping rates, as
described in section \ref{sec:symmetric}, and further restrict the
case of baths with equal fugacities, $z_l = z_r = z_0$. Assume that
these bath fugacities are subcritical, i.e., $z_0 < z_c$, where
$z_c$ is the radius of convergence of the sum (\ref{e:gc}). The
unconditioned process, with $s=0$, has a flat fugacity profile,
while any $s\neq 0$ leads to a quadratic profile
(\ref{eq:SymmetricEffectiveProfile}) whose maximum is attained at
site $k=(L+1)/2$ and equals $z_{(L+1)/2} = z_0 \cosh^2(s/2)$. Thus,
a steady state distribution exists only if $|s| <
\sqrt{\mathrm{arccosh}(2 z_c/z_0)}$, and for larger $|s|$ a
condensate forms in the middle of the system. Similar condensation
transitions occur when conditioning systems with unequal reservoir
fugacities, or with asymmetric hopping. In all these cases, the
condensates form in the bulk and their exact location is dictated by
the maximum of the fugacity profile.

We expect the supercritical phase to be composed of condensates
within the bulk which continually accumulate particles, coexisting
with a stationary ``fluid'' with a finite density. The exact pattern
of these condensates is beyond the scope of the present paper.

\subsection{Thermodynamic limit}\label{sec:TDlimit}

Consider the limit $L\to\infty$ with asymmetry $a > 1$ fixed and
define the macroscopic space variable $r=r_k$. For $r\neq 0,1$ one
obtains $y(r) = 1$ and $z(r) = \alpha \rme^{s} / (p-q+\gamma) =:
z^\ast$ independent of $r$. The stationary current of the effective
process becomes $j^\ast = (p-q)  z^\ast$. Therefore the effective
process in the thermodynamic limit is a ZRP with constant bulk
hopping rates $p,q$ as the original process, but effective left
injection rate $\alpha^\mysubtext{eff} = \alpha  \rme^{s}$. However,
there are in general boundary regions  of width $\xi=1/\log{a}$
sites with spatially decaying local driving field and associated
non-trivial exponentially decaying boundary layers of the fugacity.

In the symmetric case the localization length $\xi$ diverges. One
has $y(r) = r + \rme^{-s} (1- r)$ which leads to an algebraically
decaying local drift [see (\ref{eq:SymmetricField})]
\begin{equation}
E(r) = \frac{4 \sinh (s/2)}{L[\rme^{s/2}\,r + \rme^{-s/2}\,(1-r)]}
\end{equation}
(the $L^{-1}$ factor indicates weakly asymmetric hopping, as
discussed below). Therefore, for any $s\neq 0$ the effective ZRP has
space-dependent hopping rates. The fugacity takes the form \be z(r)
= \frac{\alpha}{\gamma} + \left( \frac{\delta}{\beta} -
\frac{\alpha}{\gamma}\right) \rme^{-s} r + (1-\rme^{-s})\left(
\frac{\delta}{\beta} - \frac{\alpha}{\gamma} \rme^{s}\right) r^2.
\ee For general $s$ this has a quadratic term which does not exist
in the ZRP with space-independent hopping rates. Even when the
fugacity profile is linear this is generated by space-dependent
hopping rates.

One may also consider the case of weakly asymmetric hopping rates,
where the asymmetry scales as $p-q = \nu/L$. Such rates correspond
to a driving field which varies substantially only on a macroscopic
length scale. Taking the $L\to\infty$ limit of \eqn
(\ref{eq:effectiveField}), and noting that in this limit $\nutilde
\to \nu/2q$, yields
\begin{equation}
E(r) = \frac{2\nu\bigl(\rme^{s+\nu} \cosh[\nu r] - \cosh[\nu(1-r)]\bigr)}
{L\bigl(\rme^{s+\nu} \sinh[\nu r] + \sinh[\nu(1-r)]\bigr)}
\end{equation}
(here the result is presented for $q = 1/2$ to simplify notation).
Other properties of weakly asymmetric systems are considered in more
detail in Section \ref{sec:MFT} below.

\subsection{Time-reversed effective dynamics}

The adjoint (time-reversed) dynamics of a process with generator
$\hat{H}$ is generally defined by $\hat{H}^\ast = \hat{P}^\ast
\hat{H}^T (\hat{P}^\ast)^{-1}$ where the diagonal matrix
$\hat{P}^\ast$ of stationary probabilities is given by $\hat{P}^\ast
\ket{\bfS} = \ket{P^\ast}$ for a stationary probability vector
$\ket{P^\ast}$ and transposed summation vector $\ket{\bfS} =
(\bra{\bfS})^T$. The system satisfies detailed balance when
$\hat{H}=\hat{H}^\ast$. This is equivalent to time-reversibility and
implies the absence of macroscopic currents. In general, however,
the absence of macroscopic currents does not imply detailed balance.
In this section we show that in our system the absence of currents
does in fact imply detailed balance.

Here, since $\hat{P}^\ast =  \hat{D} \hat{\Gamma}  $, where
$\hat{\Gamma}$ has the components of the right eigenvector of
$\hat{H}(s)$, one has \bel{adjoint} \hat{G}^\ast(s) = \hat{\Gamma}
(s) \hat{H}^T(s) \hat{\Gamma}^{-1}(s) - \epsilon_0(s). \ee Observe
that \be \hat{\Gamma}   (\hat{a}^+_k)^T \hat{\Gamma}^{-1} = x_k^{-1}
\hat{a}^-_k, \quad \hat{\Gamma}   (\hat{a}^-_k)^T \hat{\Gamma}^{-1}
= x_k \hat{a}^+_k. \ee Therefore \bea - \hat{\Gamma}   h^T_k
\hat{\Gamma}^{-1} & = & p \left(\frac{x_k}{x_{k+1}} \hat{a}_k^+
\hat{a}_{k+1}^- - \hat{d}_k\right)
+ q \left(\frac{x_{k+1}}{x_k} \hat{a}_k^- \hat{a}_{k+1}^+ - \hat{d}_{k+1}\right) \nonumber\\
 & = & q \frac{x_{k+1}}{x_k} \left( \hat{a}_k^- \hat{a}_{k+1}^+ - \hat{d}_k\right)
+ p \frac{x_k}{x_{k+1}} \left(\hat{a}_k^+ \hat{a}_{k+1}^- - \hat{d}_{k+1}\right) \\
& & + (px_k-qx_{k+1}) \left(\frac{\hat{d}_{k+1}}{x_{k+1}}  - \frac{\hat{d}_{k}}{x_{k}}\right)  \nonumber \\
 & = & q \frac{x_{k+1}}{x_k} \left( \hat{a}_k^- \hat{a}_{k+1}^+ - \hat{d}_k\right)
+ p \frac{x_k}{x_{k+1}} \left(\hat{a}_k^+ \hat{a}_{k+1}^- - \hat{d}_{k+1}\right) \nonumber \\
& & + C \left(\frac{\hat{d}_{k+1}}{x_{k+1}}  -
\frac{\hat{d}_{k}}{x_{k}}\right) \nonumber \eea and for the
boundaries \bea \fl - \hat{\Gamma}   h^T_0 \hat{\Gamma}^{-1} & = &
\alpha(x_1^{-1} \rme^{s} \hat{a}_1^-  - 1) +
\gamma( x_1 \rme^{-s} \hat{a}_1^+  - \hat{d}_1) \nonumber\\
\fl & = & \alpha x_1^{-1} \rme^{s} ( \hat{a}_1^- - \hat{d}_1) +
\gamma x_1 \rme^{-s} ( \hat{a}_1^+  - 1)
 +  (\alpha x_1^{-1} \rme^{s} - \gamma)\hat{d}_1 +  \gamma x_1  \rme^{-s}  - \alpha \\
\fl & = & \gamma x_1 \rme^{-s} ( \hat{a}_1^+  - 1)  + \alpha
x_1^{-1} \rme^{s} ( \hat{a}_1^- - \hat{d}_1) + C \left(
\frac{\hat{d}_1}{x_1} - \rme^{-s} \right). \nonumber \eea and \bea
\fl - \hat{\Gamma}   \hat{h}_L \hat{\Gamma}^{-1} & = & \delta (x_L^{-1} \hat{a}_L^- - 1)+\beta(x_L \hat{a}^+_L -\hat{d}_L) \nonumber\\
\fl & = & \beta x_L (\hat{a}_L^+ - 1)+\delta x_L^{-1} (\hat{a}^-_L
-\hat{d}_L)
+ \beta x_L  - \delta + (\delta x_L^{-1} -\beta ) \hat{d}_L  \\
\fl & = & \beta x_L (\hat{a}_L^+ - 1)+\delta x_L^{-1} (\hat{a}^-_L
-\hat{d}_L)  + C \left( 1 - \frac{\hat{d}_L}{x_L} \right). \nonumber
\eea Therefore, by the telescopic property of the sum and
\eref{energyexpression} one gets the effective adjoint dynamics \bea
\label{effdynadjoint} \hat{G}^{(0)\ast}(s) & = & - \sum_{k=1}^{L-1}
\left[ q \frac{x_{k+1}}{x_k} \left( \hat{a}_k^- \hat{a}_{k+1}^+ -
\hat{d}_k\right)
+ p \frac{x_k}{x_{k+1}} \left(\hat{a}_k^+ \hat{a}_{k+1}^- - \hat{d}_{k+1}\right) \right] \\
& & - \left[ \gamma x_1 \rme^{-s} ( \hat{a}_1^+  - 1)  + \alpha x_1^{-1} \rme^{s} ( \hat{a}_1^- - \hat{d}_1) \right] \nonumber\\
& & - \left[ \beta x_L (\hat{a}_L^+ - 1)+\delta x_L^{-1}
(\hat{a}^-_L -\hat{d}_L)\right] \nonumber \eea This is a driven ZRP
with a spatially varying driving field $E_k = -\log{a} + 2 \log{
x_{k+1}/x_k}$. By comparing transition rates one finds detailed
balance $\hat{G}^{(0)\ast}(s) = \hat{G}^{(0)}(s)$ if and only if the
stationary current \eref{currenteffective2} vanishes. Hence, for the
conditioned model, the absence of a macroscopic current also implies
detailed balance.

\section{Atypical current events via macroscopic fluctuation theory}
\label{sec:MFT}

It is our aim in this section to relate the exact microscopic
results obtained above to the optimal profiles for a large deviation
of the current as obtained from the macroscopic fluctuation theory
(MFT) \cite{Bert15}. Since the effective dynamics does not depend on
site $k$ of the local conditioning we shall consider here global
large deviations of the global time-integrated current in the whole
lattice.

The MFT can only deal with symmetric or \emph{weakly asymmetric}
hopping where we set
\begin{equation}\label{eq:WeaklyAsymmetricRates}
p - q = \nu / L
\end{equation}
and  $\nu$ is kept fixed in the thermodynamic limit of $L \to
\infty$. In what follows we employ the notation defined in
(\ref{eq:deltamu})--(\ref{eq:z0}). To conform with conventional
notations in studies of the MFT, in this section $x$ (rather than
$r$) denotes the macroscopic location along the system.

\subsection{Fluctuating hydrodynamic description}

In the limit of $L \to \infty$, and rescaling space as $x = k / L$
($0\leq x \leq 1$) and time as $1/L^2$, the evolution of the ZRP may
be described using a fluctuating hydrodynamic equation. In this
description, the density $\rho(x,t)$ evolves according to the
continuity equation
\begin{equation}\label{eq:ContinuityEq}
\dot{\rho}(x,t) = -\frac{\partial j(x,t)}{\partial x},
\end{equation}
where $j(x,t)$ is a fluctuating current. The current is given by
\begin{equation}\label{eq:FluctuatingHydro}
j(x,t) = -D(\rho) \frac{\partial \rho}{\partial x} + \nu \sigma(\rho)
+ \sqrt{\sigma(\rho)} \xi(x,t),
\end{equation}
where $\xi$ is a standard white noise, and $D(\rho)$, $\sigma(\rho)$
are drift and diffusion coefficients. The validity of the
fluctuating hydrodynamics description is based on an assumption of
local equilibrium \cite{BodineauDerrida2005WeaklyAsym}, i.e., around
each site $k=L x$ there is ``box'' of size $1\ll \ell \ll L$ sites
which, during times which are much larger than $\ell^2$ but much
shorter than $L^2$ (in the original time units, rather than the
rescaled ones), is approximately at equilibrium with density
$\rho(x)$.
%(i.e., microscopic configurations in this box are
%approximately given by the Gibbs-Boltzman distribution with this
%density).
This assumption explains why only equilibrium and linear
response coefficients ($D$ and $\sigma$) enter the macroscopic
evolution equation.

For the weakly asymmetric ZRP defined above with $q=1/2$ and $p$ as
in (\ref{eq:WeaklyAsymmetricRates}) it is possible to show
\cite{BertiniEtal2002Long,BodineauDerrida2004Additivity} that
\begin{equation}\label{eq:ZRPdrift}
D(\rho) = \frac{1}{2}\frac{\partial z(\rho)}{\partial \rho}
\end{equation}
and
\begin{equation}\label{eq:ZRPdiffusion}
\sigma(\rho) = z(\rho),
\end{equation}
where $z(\rho)$ is found by inverting the relation
(\ref{eq:ZRPdensity}) with (\ref{e:gc}).
\begin{comment}
\begin{equation}\label{eq:ZRPFugacityDensity}
\rho(z) = \frac{\partial \log \mathcal{Z}(z)}{\partial \log z},
\end{equation}
with
\begin{equation}\label{eq:ZRPpartition}
\mathcal{Z}(z) \equiv \sum_{n=0}^{\infty}  z^n \prod_{i=1}^n u(i)^{-1}.
\end{equation}
\end{comment}

\subsection{Current large deviations}

Assume that during a long period of time $T$, an atypical mean
current $j$ was observed to flow through the system. What is the
probability for such an event to happen, and what is the most likely
density profile conditioned on such an event? These questions can be
answered in the limit of $L \to \infty$ for a general driven
diffusive system which is described by \eqns
(\ref{eq:ContinuityEq})--(\ref{eq:FluctuatingHydro}) by using the
macroscopic fluctuation theory \cite{Bert15} and the additivity
principle
\cite{BertiniEtal2002Long,BodineauDerrida2005WeaklyAsym,BodineauDerrida2004Additivity,BodineauDerrida2007Review,Derrida2007}.

The
probability to see an atypical mean current $j$ during a time window
$T$ has a large deviations form
\begin{equation}\label{eq:CurrentLDFdef}
P(j) \sim \rme^{-T F(j)}.
\end{equation}
Assuming that the most probable way to generate this large
deviations event is by a static density profile $\rho^*(x)$, the
large deviations function $F(j)$ is given by
\begin{equation}
F(j) = \min_{\rho(x)} \int_0^1\frac{[j +D(\rho) \frac{d\rho}{dx} - \nu \sigma(\rho)]^2}
{2 \sigma(\rho)}\, dx.
\end{equation}
The minimization is over all profiles which satisfy the boundary
conditions dictated by the boundary reservoirs
\begin{eqnarray}
\rho(0) &=& \rho_l \equiv \rho(z_l) \nonumber \\
\rho(1) &=& \rho_r \equiv \rho(z_r),
\end{eqnarray}
where $\rho(z)$ is given in \eqns (\ref{eq:ZRPdensity}) and
(\ref{e:gc}). The typical profile $\rho^*(x)$ is then precisely the
profile for which this minimum is achieved.

In our case, it will prove useful to describe the optimal profile
using the fugacity profile $z(x)$ rather than the density. The two
are related by their usual equilibrium relation, which for the ZRP
is given in \eqns (\ref{eq:ZRPdensity}) and (\ref{e:gc}). Using the
fluctuation-dissipation relation \cite{BodineauDerrida2007Review}
\begin{equation}
D(\rho) = \frac{1}{2}\frac{d\log z}{d\rho}\, \sigma(\rho),
\end{equation}
one can transform the current large deviations function to
\begin{equation}
F(j) = \min_{z(x)} \tilde{F}[z(x),j] \equiv \min_{z(x)} \int_0^1
\frac{[j +\frac{\sigma(z)}{2z} \frac{dz}{dx} - \nu \sigma(z)]^2}
{2 \sigma(z)} dx.
\end{equation}
The boundary conditions now read $z(0) = z_l$ and $z(1) = z_r$. The
optimal fugacity profile $z^*(x)$ is the one for which the minimum
is achieved, and the optimal density profile is $\rho^*(x) =
\rho(z^*(x))$.

A general solution of the ensuing optimization problem can be
obtained as follows
\cite{BodineauDerrida2004Additivity,BodineauDerrida2007Review}. The
functional $\tilde{F}$ can be rewritten as
\begin{equation}\label{eq:OptimizationFunctional}
\tilde{F}[z(x),j] = \int_0^1 \Bigl[W(z) + V(z) \Bigl(\frac{dz}{dx}\Bigr)^2\Bigr] dx +
\int_{z_l}^{z_r} \frac{j-\nu \sigma(z')}{2z'} dz' - j\nu,
\end{equation}
where
\begin{equation}\label{eq:OptimizationAB}
V(z) = \frac{\sigma(z)}{8z^2}, \qquad \mytext{and}
\qquad W(z) = \frac{j^2}{2\sigma(z)} + \frac{\nu^2 \sigma(z)}{2}.
\end{equation}
Performing the functional derivative of $\tilde{F}$ yields
\begin{equation}
0 = \frac{\delta \tilde{F}}{\delta{z}} = \frac{dW}{dz} - \frac{dV}{dz}\Bigl(\frac{dz}{dx}\Bigr)^2
- 2 V(z)\frac{d^2 z}{dx^2}.
\end{equation}
Multiplying this equation by $dz/dx$ one obtains
\begin{equation}
\frac{d}{dx}\Bigl[W(z) - V(z)\Bigl(\frac{dz}{dx}\Bigr)^2\Bigr] = 0,
\end{equation}
which upon integrating yields the differential equation
\begin{equation}\label{eq:OptimizationODE}
\frac{dz}{dx} = \pm \sqrt{\frac{W(z)+\tilde{K}}{V(z)}} =
\pm \frac{2jz}{\sigma(z)}\sqrt{\Bigl(1+\frac{\nu \sigma(z)}{j}\Bigr)^2+K\sigma(z)}.
\end{equation}
Here $K$ and $\tilde{K}$ are integration constants which are related
by $\tilde{K} = j^2 K /2 + j\nu$. They are determined by the
boundary condition $z(1) = z_r$. Using (\ref{eq:OptimizationODE})
one may simplify a bit expression (\ref{eq:OptimizationFunctional})
evaluated at the optimal profile:
\begin{equation}\label{eq:OptimizationF}
F(j) = 2\int_0^1 W(z^*) dx + \int_{z_l}^{z_r} \frac{j-\nu \sigma(z')}{2z'} dz' +\frac{j^2 K}{2}.
\end{equation}

\subsection{Optimal profile for the ZRP}

Substituting the drift and diffusion coefficients
(\ref{eq:ZRPdrift}) and (\ref{eq:ZRPdiffusion}) in
(\ref{eq:OptimizationODE}) yields the ordinary differential equation
\begin{equation}\label{eq:ZRPODE}
\frac{dz}{dx} = \pm 2j \sqrt{\Bigl(1+\frac{\nu z}{j}\Bigr)^2+K z}.
\end{equation}
As in the microscopic calculation, the fugacity profile is
independent of the rates $u(n)$, while the density profile depends
on the form of $u(n)$ through (\ref{eq:ZRPdensity}) and
(\ref{e:gc}).

The solution of (\ref{eq:ZRPODE}) with the boundary conditions $z(0)
= z_l$ and $z(1) = z_r$ is
\begin{equation}\label{eq:OptimalProfileJ}
\fl z^*(x) = z_0 \frac{\rme^{\frac{  \Delta \mu}{2}} \sinh^2[\nu(1-x)] + \rme^{-\frac{  \Delta \mu} {2}} \sinh^2[\nu x]
+ 2 Q \sinh[\nu x]\sinh[\nu(1-x)]}{\sinh^2\nu},
\end{equation}
with
\begin{equation}\label{eq:QofJ}
Q \equiv \sqrt{1+\Bigl( \frac{j \sinh \nu}{z_0 \nu}\Bigr)^2}.
\end{equation}
When integrating \eqn (\ref{eq:ZRPODE}), the boundary condition
$z(1) = z_r$ was used to find that
\begin{equation}\label{eq:ZRPk}
K = -\frac{2\nu}{j}\Bigl[1 + \frac{z_0 \nu}{j \sinh \nu}\Bigl(Q \coth \nu -
\frac{\cosh (  \Delta \mu /2)}{\sinh \nu} \Bigr)\Bigr].
\end{equation}

An interesting limit is that of a symmetric ZRP, in which $\nu \to
0$. Taking this limit in \eqn (\ref{eq:OptimalProfileJ}) yields the
simpler quadratic expression
\begin{equation}
z^*_{\nu = 0}(x) = z_0 \bigl[e^{\frac{  \Delta \mu}{2}} (1-x)^2 + \rme^{-\frac{  \Delta \mu} {2}} x^2
+ 2 x (1-x) \sqrt{1 + (j/z_0)^2}\bigr].
\end{equation}
This expression simplifies further in the equilibrium case where
$z_l = z_r$, i.e., ${  \Delta \mu = 0}$:
\begin{equation}
z^*_{\nu,  \Delta \mu = 0}(x) = z_0 \bigl[1 +
2 x (1-x) \bigl(\sqrt{1 + (j/z_0)^2}-1\bigr)\bigr].
\end{equation}

\begin{comment}
A few fugacity profiles with atypical values of the current for
different choices of the parameters, including symmetric and
asymmetric boundary conditions and bulk dynamics, are presented in
figure \ref{fig:AtypicalZRP}. Note that the height of the fugacity
profile increases with the atypical value of $j$, and the maximal
fugacity along this profile can attain arbitrarily high values.
\end{comment}

Note that according to (\ref{eq:OptimalProfileJ}) the fugacity
profile (and hence the density profile) is symmetric around $x =
1/2$ whenever $z_l = z_r$, or equivalently $  \Delta \mu = 0$, even
when the bulk dynamics is asymmetric (i.e., when $\nu \neq 0$).

\begin{comment}
\begin{figure}
  \center
  \includegraphics[width = 0.99\textwidth]{AtypicalZRP}
  \caption{\label{fig:AtypicalZRP}The fugacity profile in a ZRP
  with atypical current: (a) $z_l=z_r$ and $\nu = 0$, (b) $z_l > z_r$ and $\nu = 0$,
  (c) $z_l=z_r$ and $\nu = 1$, (d) $z_l > z_r$ and $\nu = 1$. In all plots, the
  blue curve corresponds to the typical current [(a) $j_\text{typical} = 0$,
  (b) $j_\text{typical} = 0.2$, (c) $j_\text{typical} = 0.4$, (d)
  $j_\text{typical} \approx 0.862$]. The other curves correspond to atypical
  currents (increasing from bottom curve to top curve): $j=0$ (purple), $1.5$ (yellow), and $2$ (green) [note that
  in (a), $j=0$
  is the typical current].
  }
\end{figure}
\end{comment}

\subsection{Optimal profile in the $s$-ensemble}

As discussed above, rather than considering a system conditioned on
a specific value of the current, one may move to an ensemble in
which the current is allowed to fluctuate but is ``tilted'' towards
this atypical value. Formally, this is done by considering the
cumulant generating function of the current,(i.e., its log-Laplace
transform) $G(s)$, defined as
\begin{equation}
\rme^{T G(s)} \equiv \bigl\langle \rme^{s T j} \bigr \rangle \sim \int dj\, \rme^{T[sj-F(j)]},
\end{equation}
where (\ref{eq:CurrentLDFdef}) was used. The last integral is
dominated, in the large $T$ limit, by its saddle point, and
therefore $G(s)$ is the Legendre transform
\begin{equation}\label{eq:LegendreTransfrom}
G(s) = \max_j [sj-F(j)].
\end{equation}

In this grandcanonically conditioned ``$s$-ensemble'', the optimal
profile for a given value of $s$ is exactly the optimal profile for
the value $j(s)$ at which the maximum in \eqn
(\ref{eq:LegendreTransfrom}) is attained (this is nothing but the
equivalence of the two ensembles). This value can be found by
inverting the relation
\begin{equation}\label{eq:SofJ}
s(j) = F'(j).
\end{equation}
We shall now carry out this computation for the ZRP. Note that the
current $j$ enters the optimal profile (\ref{eq:OptimalProfileJ})
only through $Q$ given by (\ref{eq:QofJ}), and therefore obtaining
$Q(s)$ is goal of the current calculation.

The first step is to find $F(j)$, by substituting the optimal
profile in (\ref{eq:OptimizationF}). Carrying out the integrals, one
finds
\begin{equation}\label{eq:IntZ}
\fl \int_0^1 z^*(x) dx = \frac{z_0}{\nu \sinh \nu}\Bigl[Q(\nu \coth \nu - 1)
+\cosh(  \Delta \mu/2)\Bigl(\cosh \nu - \frac{\nu}{\sinh \nu}\Bigr)\Bigr],
\end{equation}
and
\begin{equation}\label{eq:IntInvZ}
\int_0^1 \frac{dz}{z^*(x)} = \frac{1}{j}\log[Q + \sqrt{Q^2 -1}].
\end{equation}
For the ZRP, one also obtains
\begin{eqnarray}\label{eq:IntZRP}
\fl \int_{z_l}^{z_r} \frac{j-\nu \sigma(z')}{2z'} dz' = \int_{z_l}^{z_r}
\frac{jdz'}{2z'} + \frac{\nu (z_l-z_r)}{2} = \\
 = \frac{\nu(z_l-z_r)}{2}
- \frac{j}{2}\log\Bigl(\frac{z_l}{z_r}\Bigr)  = z_0 \nu \sinh
\Bigl(\frac{  \Delta \mu}{2}\Bigr) - j \frac{  \Delta \mu}{2}.
\end{eqnarray}
Combining \eqns (\ref{eq:OptimizationAB}), (\ref{eq:OptimizationF}),
(\ref{eq:ZRPk}) and (\ref{eq:IntZ})--(\ref{eq:IntZRP}) yields
\begin{equation}
F(j) = \frac{z_0 \nu}{\sinh \nu} \Bigl[\cosh\Bigl(\frac{  \Delta \mu}{2} + \nu \Bigr)
- Q - \sqrt{Q^2-1}\log\bigl(\sqrt{Q^2-1}+Q\bigr)\Bigr].
\end{equation}
Then, using (\ref{eq:SofJ}), we find
\begin{equation}
\label{curr}
j(s) = \frac{z_0 \nu}{\sinh \nu}\sinh\Bigl(s +\frac{  \Delta \mu}{2} + \nu\Bigr).
\end{equation}
Finally, substituting in (\ref{eq:QofJ}) yields the simple
expression
\begin{equation}\label{eq:QofS}
Q = \cosh\Bigl(s +\frac{  \Delta \mu}{2} + \nu\Bigr),
\end{equation}
which together with (\ref{eq:OptimalProfileJ}) yields the optimal
fugacity profile for a given value of $s$. As before, it is
interesting to consider the symmetric limit of $\nu = 0$. In this
case,
\begin{equation}\label{eq:OptimalSymmetricS}
z^*_{\nu = 0}(x) = z_0 \bigl[\rme^{\frac{  \Delta \mu}{2}} (1-x)^2 + \rme^{-\frac{  \Delta \mu} {2}} x^2
+ 2 \cosh\Bigl(s + \frac{  \Delta \mu}{2}\Bigr) x (1-x) \bigr].
\end{equation}

\subsection{Comparison with the microscopic calculation}

As discussed in Section \ref{sec:TDlimit}, in the weakly asymmetric
limit of (\ref{eq:WeaklyAsymmetricRates}) with $q=1/2$, and when
$L\to\infty$, one has $\nutilde \to \nu$ [where $\nutilde$ is the
asymmetry parameter defined in (\ref{eq:nutilde})]. Therefore, in
this limit the microscopic fugacity profile
(\ref{eq:OptimalProfileJmicroscopic})--(\ref{eq:Qtilde}) is
precisely the same as the macroscopic expressions
(\ref{eq:OptimalProfileJ}) and (\ref{eq:QofS}). Similarly, the
macroscopic expression \eref{curr} for the current in the
$s$-ensemble coincides in this limit with exact expression
\eref{currbf} obtained for barrier-free reservoir coupling with weak
asymmetry. Indeed, figures \ref{f:WeakAsym} and \ref{f:sym} present
fugacity profiles for symmetric and weakly-asymmetric hopping and
several values of $s$, and demonstrate the agreement between the
microscopic and macroscopic results.

An interesting observation is that the microscopic expression
(\ref{eq:OptimalProfileJmicroscopic}) for the optimal profile and
macroscopic one (\ref{eq:OptimalProfileJ}) have the same functional
form \emph{even when $L$ is small and the asymmetry is not weak},
provided one uses the definition $x=k/(L+1)$.  In particular, by
substituting $\nu = c L$ in the macroscopic profile
(\ref{eq:OptimalProfileJ}) (with a constant $c$) it is possible to
obtain the correct profile for strongly asymmetric hopping (i.e.,
with $p \neq q$ independent of $L$). However, using the macroscopic
theory beyond its limit of validity comes with a price: The precise
value of the prefactor $c$ which corresponds to given values of $p$
and $q$ cannot be determined within the macroscopic approach, and
only the microscopic one yields equation (\ref{eq:nutilde}).

\begin{comment}
\begin{multline} z^*(k) =
\frac{z_0}{(v^2-w^2)^2}\Bigl[(v^2-w^2)\cosh\Bigl(\frac{s}{2}\Bigr) +
\Bigl(v^2 + w^2 - 2 v w
a^{\frac{L+1}{2}-k}\Bigr)\sinh\Bigl(\frac{s}{2}\Bigr) \Bigr]\times \\
\Bigl[(v^2-w^2)\cosh\Bigl(\frac{s+  \Delta \mu}{2}\Bigr) + \Bigl(v^2 +
w^2 - 2 v w a^{k-\frac{L+1}{2}}\Bigr)\sinh\Bigl(\frac{s+  \Delta
\mu}{2}\Bigr) \Bigr],
\end{multline}
where $v \equiv a^{L/2}p$, $w \equiv a^{1/2}q$ and $a \equiv p/q$.
Substituting (\ref{eq:WeaklyAsymmetricRates}), $p=1/2$ and $k=L x$
and taking the limit $L \to \infty$ leads to the macroscopic result
(\ref{eq:OptimalProfileJ}) and (\ref{eq:QofS}).
\end{comment}

%\section{Conclusion}
%
%
%In these notes we have presented a macroscopic calculation (based on
%the macroscopic fluctuation theory and the additivity principle) of
%the typical profile of a ZRP with open boundaries conditioned on
%having an atypical current. The macroscopic results obtained here
%are the same as the large $L$ limit of the profile obtained from the
%microscopic calculation.

%\section{Final remarks}

%\section*{Acknowledgements}
\ack
The authors thanks D. Gabrielli, B. Meerson, and P.L. Krapivsky
for inspiring discussions and the Galileo Galilei Institute for
Theoretical Physics for hospitality. Partial support by the INFN
during the completion of this work is gratefully acknowledged. The
support of the Israel Science Foundation  (ISF) and of the Minerva
Foundation with funding from the Federal German Ministry for
Education and Research is gratefully acknowledged.

\bigskip

\bigskip

%\appendix

%\section{Vector spaces, inner product and tensor product}
%\label{tensor}

\bibliographystyle{hunsrtnat}
\bibliography{ZRP_atypical}

\end{document}